\documentclass{aa}  

\usepackage{graphicx}
\usepackage{txfonts}
\usepackage{url}

\usepackage{subfig}

\usepackage{natbib,twoopt}
\usepackage[breaklinks=true]{hyperref} %% to avoid \citeads line fills
\usepackage{pdflscape}

\begin{document}

   \title{ Kepler-411: a four-planet system with an active host star \thanks{Tables 2 , 3 and 4 are available at the CDS via anonymous ftp to cdsarc.u-strasbg.fr (130.79.128.5) or via    http://cdsarc.u-strasbg.fr/viz-bin/qcat?J/A+A/000/A00}}

  % \subtitle{I. Overviewing the $\kappa$-mechanism}

   \author{Leilei  Sun\inst{1,2,3}
          \and
         P. Ioannidis\inst{4}
          \and
           Shenghong  Gu\inst{1,2,3}
           \and 
           J. H. M. M. Schmitt\inst{4} 
          \and
           Xiaobin  Wang\inst{1,2,3}
          \and
           M.B.N.  Kouwenhoven\inst{5}    
          }

   \institute{Yunnan Observatories, Chinese Academy of Sciences, Kunming 650216, China
%              \email{sunleilei@ynao.ac.cn,wangxb@ynao.ac.cn}
       \and 
              Key Laboratory for the Structure and Evolution of Celestial Objects, Chinese Academy of Sciences, Kunming 650216, China
        \and 
               University of Chinese Academy of Sciences, Beijing 100049, China
        \and
            Hamburger Sternwarte, Universität Hamburg, Gojenbergsweg 112, 21029 Hamburg, Germany
%             \email{pioannidis@hs.uni-hamburg.de,st8h101@hs.uni-hamburg.de}
         \and
            Department of Mathematical Sciences, Xian Jiaotong-Liverpool University (XJTLU), 111 Renai Rd., Suzhou Dushu Lake Science and Education Innovation District, Suzhou Industrial Park, Suzhou 215123, China\\
             \fnmsep \thanks{Send offprint requests to: X. Wang, L. Sun, P. Ioannidis, e-mails:\textbf { wangxb@ynao.ac.cn, sunleilei@ynao.ac.cn, pioannidis@hs.uni-hamburg.de } }
             }

   \date{Received  ..., ...; accepted  ..., ...}

 \abstract{We present a detailed characterization of the Kepler-411 system (KOI 1781). This system was previously known to host two transiting planets: one with a period of 3 days ($R=2.4R_\oplus$; Kepler-411b) and one with a period of 7.8 days ($R=4.4R_\oplus$; Kepler-411c), as well as a transiting planetary candidate with a   58-day period ($R=3.3R_\oplus$; KOI 1781.03) from Kepler photometry. Here, we combine Kepler photometry data and new transit timing variation (TTV) measurements from all the Kepler quarters with previous adaptive-optics imaging results, and dynamical simulations, in order to constrain the properties of the Kepler-411 system. From our analysis, we obtain masses of 25.6$\pm$2.6$M_\oplus$ for Kepler-411b and 26.4$\pm$5.9 $M_\oplus$ for Kepler-411c, and we confirm the planetary nature of KOI 1781.03 with a mass of 15.2$\pm$5.1$M_\oplus$, hence the name Kepler-411d. Furthermore, by assuming near-coplanarity of the system (mutual inclination below $30^\circ$), we discover a nontransiting planet, Kepler-411e, with a mass of 10.8$\pm$1.1$M_\oplus$ on a 31.5-day orbit, which has a strong dynamical interaction with Kepler-411d. With densities of $1.71\pm0.39$~g\,cm$^{-3}$ and $2.32\pm0.83$~g\,cm$^{-3}$, both Kepler-411c and Kepler-411d belong to the group of planets with a massive core and a significant fraction of volatiles. Although Kepler-411b has a sub-Neptune size, it belongs to the group of rocky planets.}


   \maketitle
%
%-------------------------------------------------------------------
\section{Introduction}

The data from the {\it Kepler} mission have revealed thousands of exoplanets and exoplanet candidates through the transit method~\citep{Boru2010}. After the demise of the ``old'', primary Kepler mission, the extended Kepler mission K2 has been revealing many new exoplanet candidates \citep{Howe2014}. While being very powerful, the transit method
cannot provide the masses of the discovered transiting bodies, and therefore mass measurements have been carried out for only a rather small fraction of the discovered planetary candidates, either with the radial velocity (RV) method or with the transit timing variation (TTV) method~\citep{Jont2015,Hadd2017}. The RV technique measures the reflex motion of a host star induced by its planet, and hence the signal shrinks with increasing orbital distance, while the latter detects the gravitational perturbations between planets in multi-planet systems by precisely measuring transit times and subsequently fitting dynamical models (or derived analytic formulae) to the observed  TTVs~\citep{Mira2002,Agol2005,Holm2005,Holm2010,Nesv2012,Lith2012,Xie2014,Jont2015,Deck2015,Agol2016}. Hence, the TTV technique is very sensitive to gravitational interactions between planets that are closely spaced or near orbital resonances, in addition to the positive dependence on the orbital distance~\citep{Agol2005,Holm2005}. Characterizing planets using TTVs is therefore complementary to the RV method~\citep{Agol2005,Holm2005,Nesv2008,Nesv2012,Jont2015,Agol2017}.

Through an analysis of the continuous, high-precision photometric data collected by the {\it Kepler} space telescope, one can not only determine the radii and orbital paramaters of transiting exoplanets, but can also infer, or at least constrain, the masses of the exoplanets by studying their dynamical interactions. The planetary mass is a fundamental  physical parameter of an exoplanet, which is required for the investigation of further planetary properties, such as the composition, structure, and also the formation and evolution of the planetary system~\citep{Seag2007,Fort2007,Bara2008,Bara2010,Enoc2012,Lope2012,Lope2014,Kouw2016}. 

The vast majority of the host stars of the currently known exoplanets are of late spectral type and are therefore to some degree magnetically active. Specifically, stellar activity in late-type stars induces photometric modulations and apparent RV variations, which may hamper the detection of planets and the accurate measurement of the system parameters~\citep{Sozz2007,Enoc2010,Czes2009,Osha2015}. Star spots can deform the transit light curves, for example through spot-crossing events, and can thus possibly mimic the TTV signals induced by the gravitational perturbations of other planets~\citep{Osha2013b,Daas2014,Ioan2014,Ioan2016a,Ioan2016b,Sun2017}. Therefore, appropriately accounting for the effects of stellar activity on transit light curves becomes absolutely necessary for characterizing exoplanets orbiting relatively active host stars.  

In this study we present our comprehensive analysis and characterization of the multiple planetary system Kepler-411. In Section~\ref{sec_target} we provide a brief description of our target system and in Section~3 we describe the methods used to extract the transit signals and derive the planetary properties. In Section~4, we give the results of our TTVs and dynamical simulation analyses and in Section~5, we discuss the new results and constrain the compositions of Kepler-411's transiting planets. Finally, we summarize the study in Section~6. 

\section{The Kepler-411 planetary system}

\label{sec_target}

The host star, Kepler-411, is a K2V-type star located at a distance of $153.59\pm0.48$~pc~\citep{Gaia2018,Lind2018,Berg2018} exhibiting features that indicate relatively strong magnetic activity. Besides rotational modulation with a periodicity of about 10 days as seen in the {\it Kepler} light curve \citep{Hira2014}, we find that strong emissions are detected in the cores of the $Ca$ $II$ $H\&K$ lines by visually inspecting the spectroscopic data in the Keck Observatory Archive\footnote{https://www2.keck.hawaii.edu/koa/public/koa.php}.  Kepler-411 has a red companion at an angular separation of $(3.37\pm 0.05)''$, which is fainter than Kepler-411 by $\Delta J= 2.71 \pm 0.1 $ mag~\citep{Wang2014}. 

Kepler-411 was previously identified as KOI~1781. Three planetary candidates around KOI~1781 were successively discovered by the {\it Kepler} transiting planet search pipeline and the public Planet Hunters project~\citep{Tene2012,Schw2012,Bata2013,Rowe2014};  TTV signals of these three planetary candidates were also detected~\citep{Ford2012, Maze2013,Rowe2014}. Subsequently, the planetary nature of KOI 1781.02 ($p_{orb}=3.0$ days; Kepler-411b) and KOI 1781.01 ($p_{orb}=7.8$ days; Kepler-411c) were confirmed by~\citet{Wang2014} and~\citet{Mort2016}, respectively, hence the designation, Kepler-411. \citet{Wang2014} and~\cite{Mort2016} derived a confidence level for the planetary nature of KOI 1781.03 ($p_{orb}=58$ days; hereafter, we designate this body as Kepler-411d) of over 98.7\%, but the planetary nature of this object has not yet be confirmed.

Most interestingly, the amplitude of the TTVs seen in Kepler-411d reaches up to about 50~minutes, while those of two confirmed planets are smaller than 5~minutes. Furthermore, the three orbital periods do not approach any commensurability, that is, mean-motion resonances whereby the perturbations between planets are amplified and therefore induce significant TTV signals. To further unravel the planetary architecture of the Kepler-411 system, we continue our investigation of this peculiar planetary system, and present our analysis below.

%--------------------------------------------------------------------
\section{Data analysis}

\label{sec_dataanalysis}

\subsection{Data preparation and transit searching}

\label{data_preparation}

\begin{figure*}
   \centering
   \includegraphics[width=19.35cm]{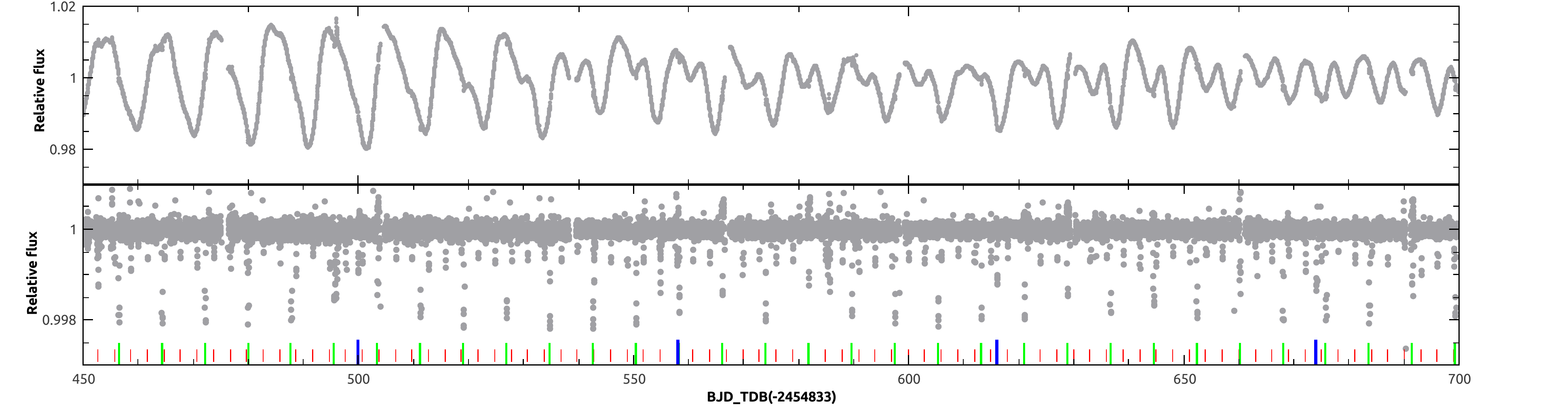}
   \caption{Part of the long-cadence data of Kepler-411, with apparent rotational modulations (upper panel). The lower panel shows the light curves detrended using the Savitzky-Golay filter. The red, green, and blue vertical lines in the lower panel mark the transit events of three transiting planets of Kepler-411.}
              \label{FigGam1}%
    \end{figure*}

The {\it Kepler} data of Kepler-411 were downloaded from the MAST archive\footnote{https://archive.stsci.edu/kepler}, which contains the data recorded in the 17 quarters; we use both long- and short-cadence data  (available for quarters Q10 to Q17). We use the PDCSAP data for our analysis. We show a part of the light curve of Kepler-411 in the upper panel of Fig.~\ref{FigGam1}. One easily identifies peak-to-peak variations of $2-4\%$ in the light curve, indicating that the host star of Kepler-411 is relatively active. Through an analysis of the light curve with the generalized Lomb-Scargle periodogram~\citep{Zech2009}, we find a clear rotation period of $10.4\pm 0.03$ days. 

To characterize the system and to search for and analyze TTVs, it is essential to remove the effects of stellar activity as much as possible.  To that end we adopt the same methodology as in~\citet{Ioan2014} to extract all transit signals from {\it Kepler} data and detrend each transit light curve.  For ``normal'' transit signals, that is, single-planet transits, we select a window of twice the transit duration, approximately centered on the mid-transit time of each transit light curve, and detrend each light curve with a second-order polynomial fit to the out-of-transit data. 

For ``anomalous'' transits, that is, two or more planet  transits, we use a wider window (i.e., a window of  triple the transit duration of Kepler-411d) to retrieve those transits where two or more planets transit the host star simultaneously. Furthermore, to visually validate our normalization of the transit light curves, we use a Savitzky-Golay filter set to second order with a 16-hour window to remove the rotational modulations in the light curve of Kepler-411. Prior to filtering the light curves, we substitute all transits with the polynomial fit results of the out-of-transit data in order to retain the characteristics of each transit extracted with our procedure. Through inspection of the detrended light curve, we can easily identify inappropriately detrended transit signals and repeat the detrending procedure for these. Inappropriately detrended transit signals usually originate from incorrect fits due to outliers in out-of-transit data; therefore we fit the out-of-transit again without these outliers. To demonstrate our methodology, we show a part of the Kepler-411 light curve with rotational modulation signals filtered out in the bottom panel of Fig.~\ref{FigGam1}.

As described above, stellar activity may hide the transit signals of Earth-like exoplanets. In order to search for possible transit signals which are obscured by stellar activity, we employ the Box-fitting Least Square algorithm~\citep{Kova2002} to search the  light curves detrended by using the Savitzky-Golay filter with the known transit signals removed. We find no evidence for any significant transit-like signal. 

\subsection{Model fitting}
\label{sec_modeling}
Stellar activity can not only create bumps on a transit light curve when star spots (or faculae) are occulted by the planet, but can also lead to variable depths of normalized transit light curves due to the influence of stellar activity on the total flux of the host star~\citep{Czes2009,Ioan2016b,Sun2017}. 

We employ the Spot and Transit Modeling Tool (STMT), as described in~\citet{Sun2017}, to model the normal transit light curves of Kepler-411. The recently updated STMT has not only the capability to model the combined effect of spots and planets on the light curve simultaneously, but can also be used to efficiently simulate double transit curves of an active host star using the analytic formulae derived by~\citet{Pal2012} and~\citet{Mont2014}. Even light curves that include rare mutual planet-planet eclipse events can be modeled by the updated STMT. The new version of the STMT code was successfully tested on the system KOI-94 with mutually eclipsing planets KOI-94d and KOI-94e~\citep{Hira2012}. 

We adopt the standard quadratic limb-darkening law, interpolating the limb darkening coefficients from the table of~\citet{Clar2013} based on the stellar parameters of Kepler-411 derived by~\citet{Wang2014}. Because of the low signal-to-noise ratio (S/N) of the transits sampled with short-cadence data, we use the mean density of the host star, $\rho_{*}$, derived from interpolating the empirical relation of~\citet{Torr2010} as a prior to fit these data. We note that the mean density of the host star can be inferred from normalized orbital semi-major axis $a/R$ combined with Kepler's third law when modeling the transit light curves. Furthermore, we adopt the calibration derived by~\citet{Enoc2012} to convert the mean density $\rho_{*}$, the effective temperature $T_{\rm eff}$, and the metallicity $[Fe/H]$ of the host star into its mass, a method which is also used by the SuperWASP team and other groups~\citep{Came2006,Came2007,Brow2012,Wang2014a,Sun2015}. Therefore, the ratios between the radii and masses of the exoplanet and the host star, respectively, can be converted into the actual planetary radius and mass, and hence the uncertainty of the planet's radius and mass due to those in the host star can be appropriately estimated.  

\begin{figure}
   \centering
   \includegraphics[width=9.3cm]{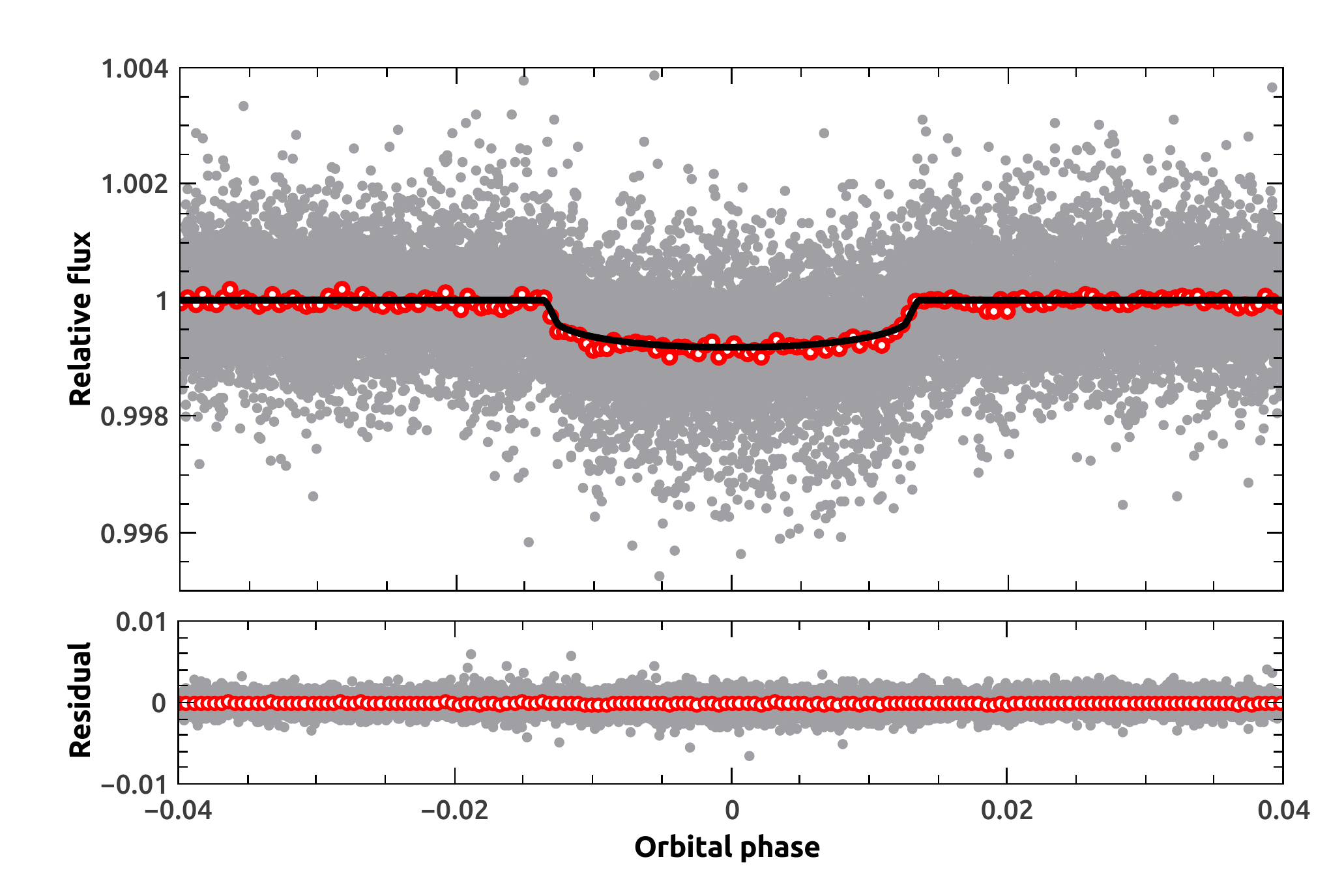}
   \caption{Fitting result of the short-cadence transit light curves of Kepler-411b.}
              \label{Figlca}%
    \end{figure}

\begin{figure}
   \centering
   \includegraphics[width=9.3cm]{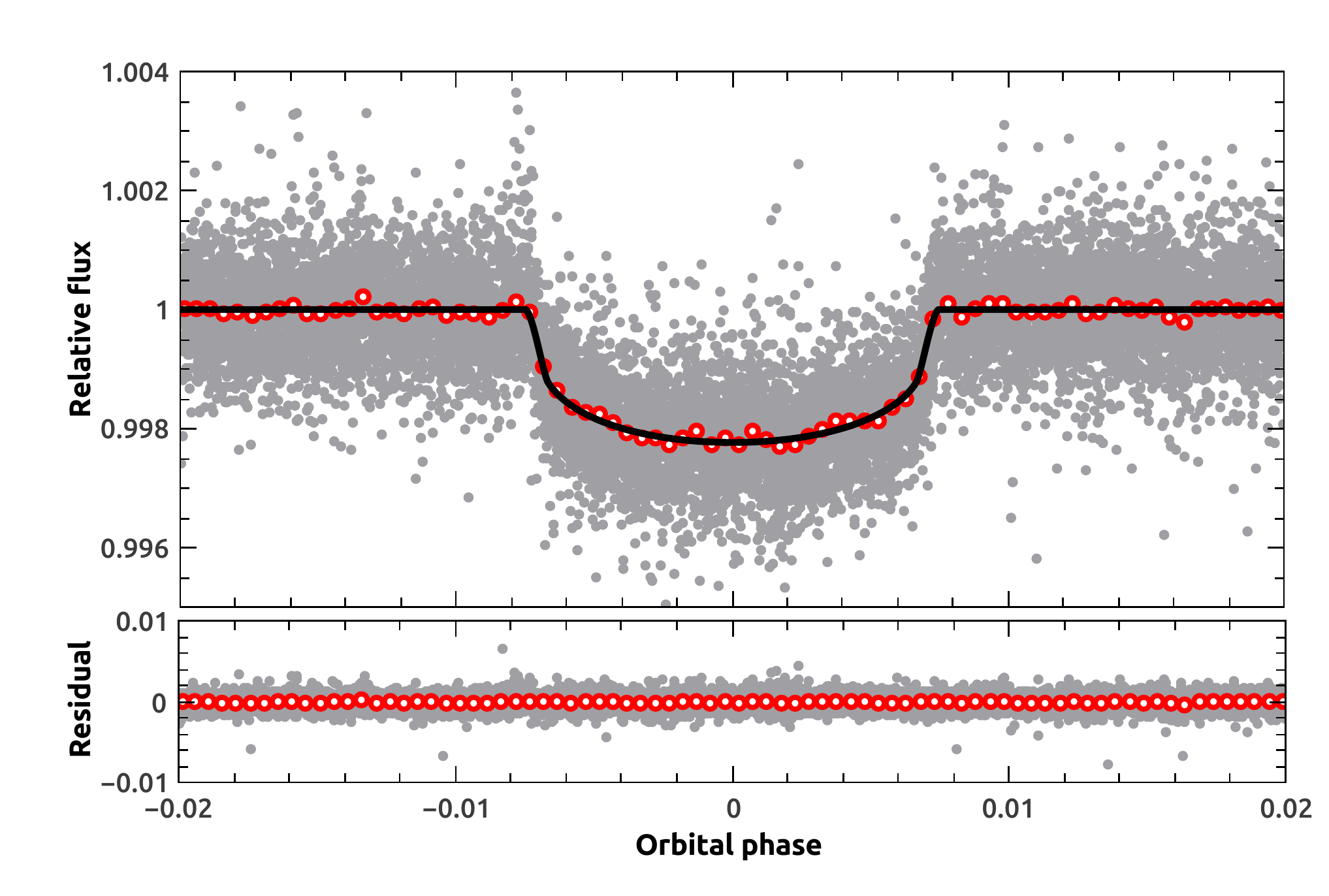}
   \caption{Fitting result of the short-cadence transit light curves of Kepler-411c.}
              \label{Figlcb}%
    \end{figure}

\begin{figure}
   \centering
   \includegraphics[width=9.3cm]{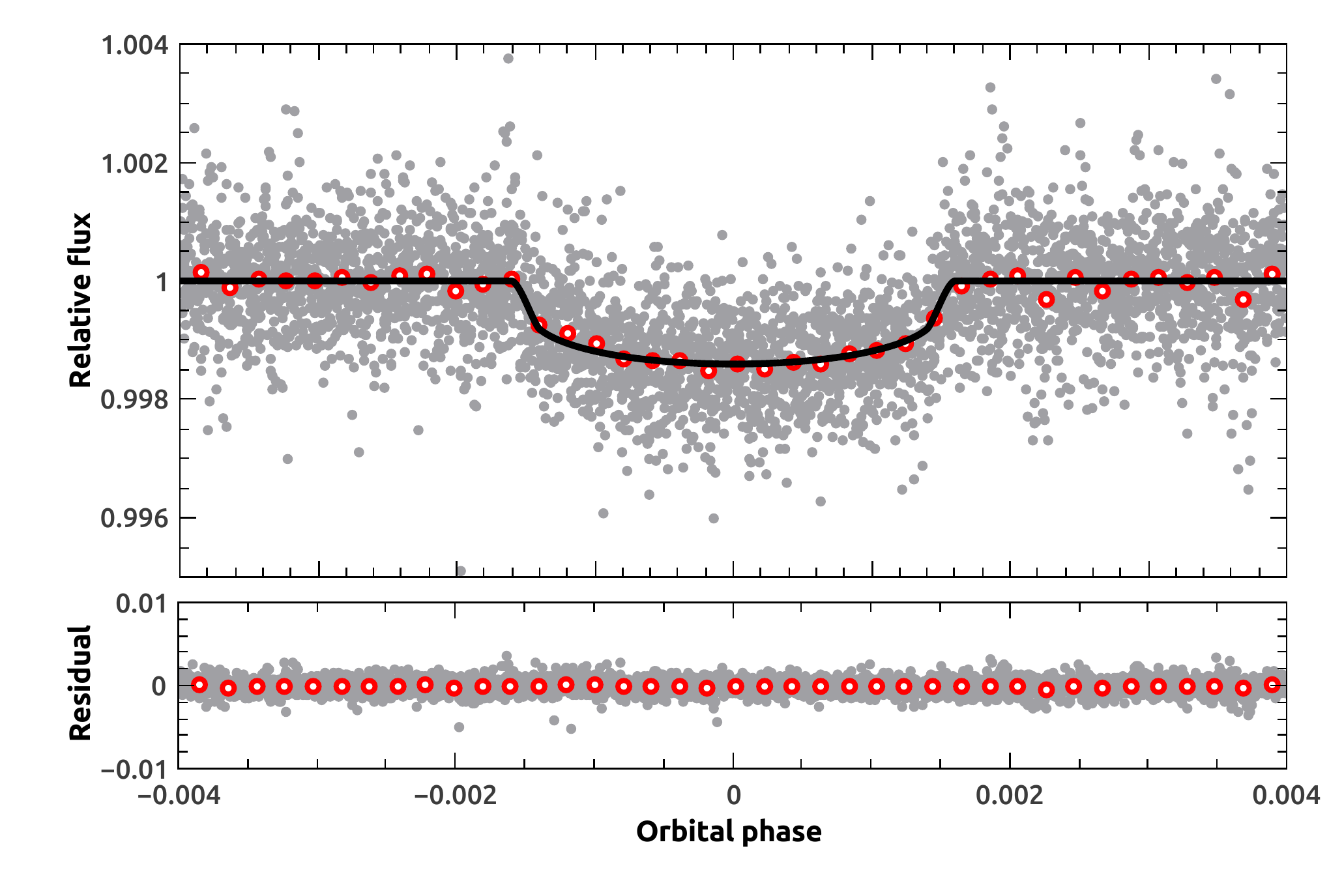}
   \caption{Fitting result of short-cadence transit light curves of Kepler-411d. }
              \label{Figlcc}%
    \end{figure}

We first use the normal transit light curves extracted from the short-cadence data to construct a fiducial model of each known transiting body. We note that we only select the transit light curves near the local flux maximum to calculate the physical parameters, since these local maxima tend to be less affected by star spots. These fiducial models form the basis for modeling spot crossing events and double-transit events, because there is a strong degeneracy between transit parameters and spot properties derived from spot-distorted transit light curves \citep{Sun2017}. The short-cadence data constrain the parameters of a transiting body better than the long-cadence data. Short-cadence data, which have a sampling interval of  $\sim 1$~minute, on the one hand, keep the morphology of the transit light curve; on the other hand, when many short-cadence light curves are modeled simultaneously, short-cadence data with sufficient sampling can possibly alleviate the influence of stellar activity and the systematic errors of the light curves of Kepler-411 on the measurement of transit parameters. In Figs.~\ref{Figlca}, \ref{Figlcb}, and \ref{Figlcc} we show our best-fit results of the respective transit light curves; we further list the best-fit system parameters of each planet in Table~\ref{table:1} as derived from modeling the transit light curves.

\begin{figure}
   \centering
   \includegraphics[width=9.3cm]{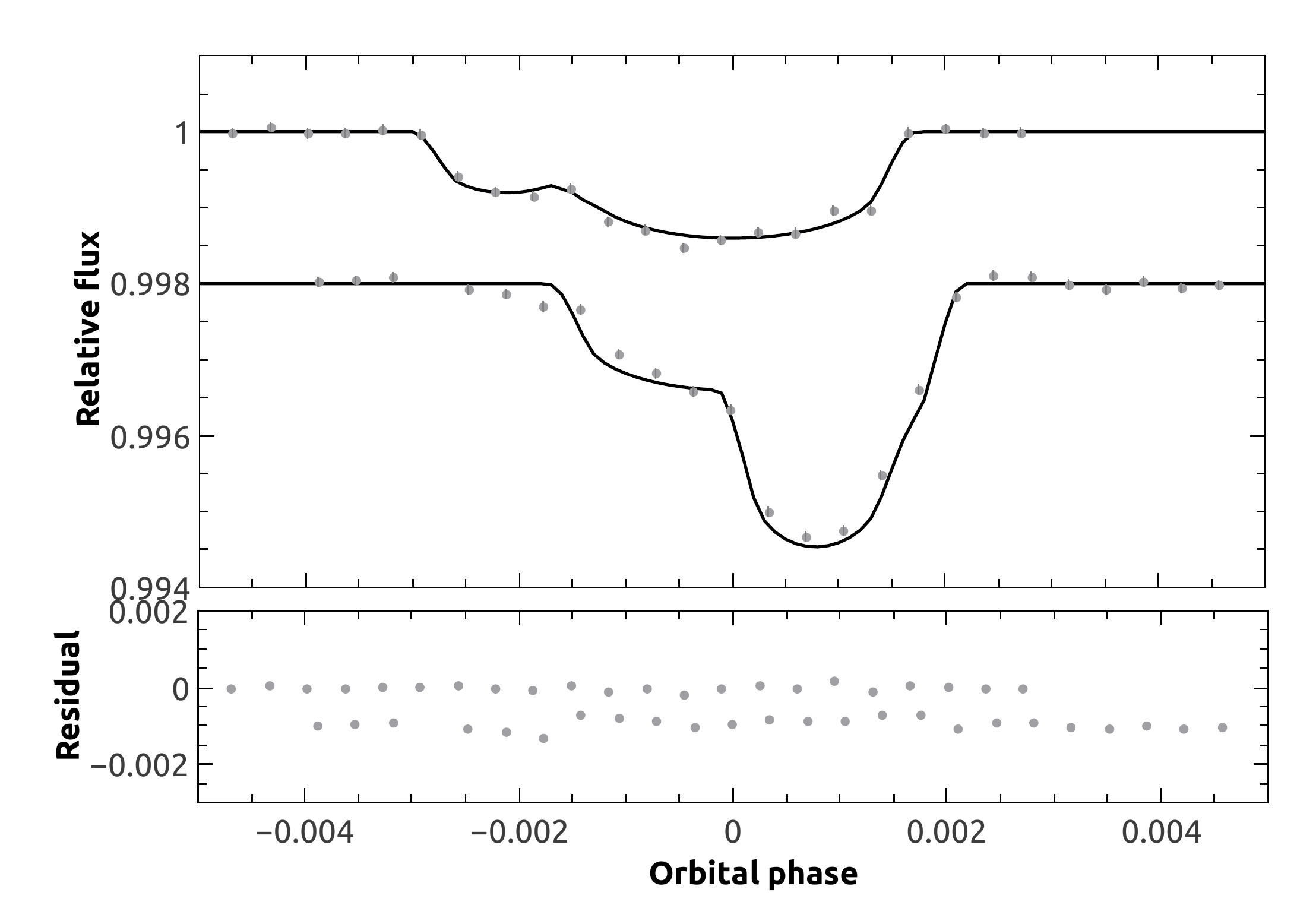}
   \caption{The first light curve shows a double transit of Kepler-411b and Kepler-411d. The second light curve shows a double transit of Kepler-411c and Kepler-411d. The residuals of the second light curve have been shifted for clarity. The physical parameters of the transit models of planets Kepler-411b, c, and d are those derived using the short-cadence data.}
              \label{double_lc}%
    \end{figure}

Subsequently, we calculate the transit times through fitting each transit light curve extracted from the long-cadence data with physical parameters fixed to the values derived from the short-cadence data. We adopt the resampling strategy proposed by~\citet{Kipp2010} to construct the model of transit light curves in the long-cadence sampling. Since only eight quarters of short-cadence data are available for the four-year observing baseline, we only use the long-cadence data to calculate the TTVs. For the transit light curves with apparent bumps in the ingress and/or egress parts, we adopt a similar strategy to~\citet{Sun2017} to model these light curves, in order to remove the possible influences on the measured transit times induced by spot-crossing events. To briefly summarize, we employ one or two circular spot(s) to reproduce the distorted light curves by spot-crossing events with the transit model fixed. We note that we do not impose the constraint that some consecutive transits are modeled with the same spots. The derived transit times are listed in Table~\ref{table:2}, Table~\ref{table:3} and Table~\ref{table:4}. We also fit the double-transit light curves with these transit models using the updated STMT code. In Fig.~\ref{double_lc} we show two double-transit events to demonstrate the validity of our methodology. We note that the fits of the double transits significantly improve the TTV measurements of Kepler-411d, since only seventeen transit events of this planet have been recorded during the four years of observations. Finally, we derive a more precise stellar mass ($M_{*}=0.870\pm0.039M_\odot$) and radius ($R_{*}=0.820\pm0.018R_\odot$) using the improved measurement of the stellar mean density obtained through our transit light-curve modeling with more data than before. These values are consistent with the previous measurements by~\citet{Wang2014}, that is, $M_*=0.83^{+0.04}_{-0.10}M_\odot$ and $R_*=0.79^{+0.07}_{-0.06}R_\odot$.

\section{Transit timing variation and dynamical stability analysis}

\subsection{Transit timing variation inversion}

\subsubsection{Transit timing variation inversion code}
\label{inversion_code}

We employ the widely used TTV inversion code TTVFast~\citep{Deck2014} to generate model transit times of each of the planets at the observed epoch. The Genetic Evolution Markov Chain (GEMC) and DiffeRential Evolution Adaptive Metropolis (DREAM) algorithms are used in the TTV inversion code to perform parameter optimization and estimation as implemented in the STMT code (see \cite{Sun2017} for further details). GEMC is an efficient optimization algorithm developed by~\citet{Treg2013}, which combines the superior global optimization power of genetic algorithms~\citep{Char1995} with the capability of parameter estimation of widely applied Differential Evolution Markov Chain Monte Carlo (DEMCMC) algorithms ~\citep{ter2006} in the exoplanet research field for simulating high-dimensional, multi-modal systems. The DREAM algorithm, proposed by~\citet{Vrug2009}, is another derivative of the DEMCMC algorithms and adopts a self-adaptive randomized subspace sampling strategy, which inherits most of the advantageous properties of the DEMCMC algorithm and performs better through increasing the diversity of the exploration of parameter space. When the code carries out the TTV inversion, it generates model TTVs and then fits them to the measured TTVs. Finally, for a preliminary evaluation of the long-term stability of the synthetic system, we implemented a Hill stability criterion in our code to remove the most unstable planetary architectures~\citep{Marc1982,Glad1993}. In addition to evaluating synthetic systems, we also tested our code on the KOI-872 system and found that the optimal solution was identical to that of~\citet{Nesv2012}.

\subsubsection{Initial TTV inversion of Kepler-411d}
\label{sec_initial}
As mentioned in the introduction, only Kepler-411d shows large TTVs, while no commensurability is found among the orbital periods of the three known transiting bodies. These observations suggest that there is (at least) one more (nonvisible) planet in orbit around Kepler-411, in addition to the three transiting planets. This additional planet may be in (or near) a mean-motion resonance with Kepler-411d. In order to test this hypothesis, we use our TTV inversion code to search for the optimal orbital architecture of this new perturber. To improve the efficiency of the dynamical modeling, we initially only include Kepler-411d in addition to the new perturber, because of the the weak dynamical interactions of Kepler-411d with the two other known planets. 

As the TTV patterns are most sensitive to the perturber's mass, the orbital period (or orbital separation), the eccentricity, and the mutual inclination of the orbit~\citep{Agol2005,Holm2005,Nesv2008,Nesv2009,Nesv2010,Xie2014,Agol2017}, we set appropriate constraints to these parameters during the search for the optimal solution. For the mass of the perturber, we impose a uniform prior between $0.1M_\oplus$ and $300M_\oplus$. In order to cover the possible architectures, the search interval for the orbital period of the  perturber is set uniformly between 3 and 580 days (the lower limit corresponds to the orbital period of the innermost planet, while the upper limit is set to ten times the orbital period of  Kepler-411d). The possibility of wider-orbit perturbers is discussed in Sects.~\ref{inclined orbit} and~\ref{Possible_influence}. We find that the inversion code cannot always find the optimal solution for synthetic TTVs, which are similar to the TTVs of Kepler-411d, when the search interval of the orbital period of the perturber is much larger than 0.5 days and trial chains of 1,000 are used in the GEMC algorithm. However, the code always finds the optimal solution when the search interval is set to 0.5~days for trial chains of 1,000 in the GEMC algorithm. Therefore, we split the search interval of the orbital period of the perturber into multiples of the 0.5~day interval and search all subintervals each time when we use our code to carry out the initial inversion of the TTVs of Kepler-411d. 

We impose a uniform prior to the eccentricity between $e=0$ and $e=0.2$ as required by the long-term stability of the system, which is demonstrated in the Kepler compact multi-planet systems~\citep{Liss2011a,Fabr2014,Xie2016}. In addition, a uniform prior is imposed on the mutual inclination, which is constrained between $0^\circ$ and $30^\circ$~\citep{Nesv2009}, mainly because larger mutual inclinations will probably trigger significant transit duration variations (TDV) for compact multiple planetary systems and tend to make this kind of system unstable for a long run~\citep{Nesv2012,Mill2017}. On the other hand, no significant TDV signals of Kepler-411d have been detected (i.e., the root mean square of the TDVs of  Kepler-411d is almost equal to the median of the uncertainties of transit duration times). The upper limit of the variation rate of the transit durations of  Kepler-411d is about 4 mins/1300 days. 

Besides these four parameters, three more parameters are initialized: the longitude of the ascending node $\Omega$, the argument of periastron $\omega$, and the mean anomaly $M$ at a reference time~\citep{Deck2014}. The tightly packed, multi-planet \textit{Kepler} systems tend to be coplanar, like our solar system~\citep{Winn2015,Xie2016}. We explicitly fix the $\Omega$ of Kepler-411 d to zero and vary the $\Omega$ of the new perturber among the interval between $-30^\circ$ and $30^\circ$, in order to improve efficiency of TTV modeling, and the search intervals of argument of periastron and mean anomaly are set to the physical boundary, that is, $\omega \in (-180^\circ$,180$^\circ$) and $M \in(0^\circ,360^\circ$). So as to accelerate the MCMC sampling, $\sqrt{e}\cos\omega$ and $\sqrt{e}\sin\omega$ are used as free parameters to explore the parameter spaces of $e$ and $\omega$ in our inversion code, where $e$ and $\omega$ are the eccentricity and argument of periastron, respectively~\citep{Ford2005,East2013}. 

We find that the dynamical modeling uniquely generates the reduced minimum $\chi^2_{\nu}$ of 3.8 (i.e., the reduced minimum $\chi^2_{\nu}$ which is larger than unity is probably induced by the underestimation of transit-time uncertainties) when the orbital period of the perturber is equal to 31.5~days. Taking the possible underestimation of transit-time uncertainties into account, we use the reduced $\chi^2_{\nu}$ of the optimal fitting derived from GEMC simulation in the TTV inversion code to scale the errors of measured transit times (namely, all errors are multiplied by a factor of $\sqrt{\chi^2_{\nu}}$), and then employ the MCMC algorithm to sample the parameter posterior distributions based on the scaled errors. The orbital period ratio between Kepler-411d and the newly identified perturber is 1.84, which is near the 2:1 (or 9:5) mean-motion resonance with Kepler-411d. Moreover, the orbital-period ratio between Kepler-411c and the perturber is 4.02. The impact parameter $b$ of the perturber is 1.6 (i.e., $b \gg 1+R_p/R_{*}$) based on the extracted orbital architecture. This provides strong evidence for the hypothesis that the 31.5-day perturber is a nontransiting planet. Hereafter, we label this new planet as Kepler-411e.

\subsubsection{Joint transit-timing-variation inversions}
\label{sec-joint}
As shown in Sect.~\ref{sec_initial}, an additional, nontransiting fourth planet Kepler-411e is necessary to reproduce the measured TTVs of Kepler-411 d. Moreover, Kepler-411 d is shown to be a genuine planetary companion of Kepler-411 (see Sect.~\ref{Possible_influence}). In this subsection, we jointly simulate the TTVs of the three known transit signals assuming that one additional nontransiting perturber exists in the system.

\begin{figure}
   \centering
   \includegraphics[width=9.5cm]{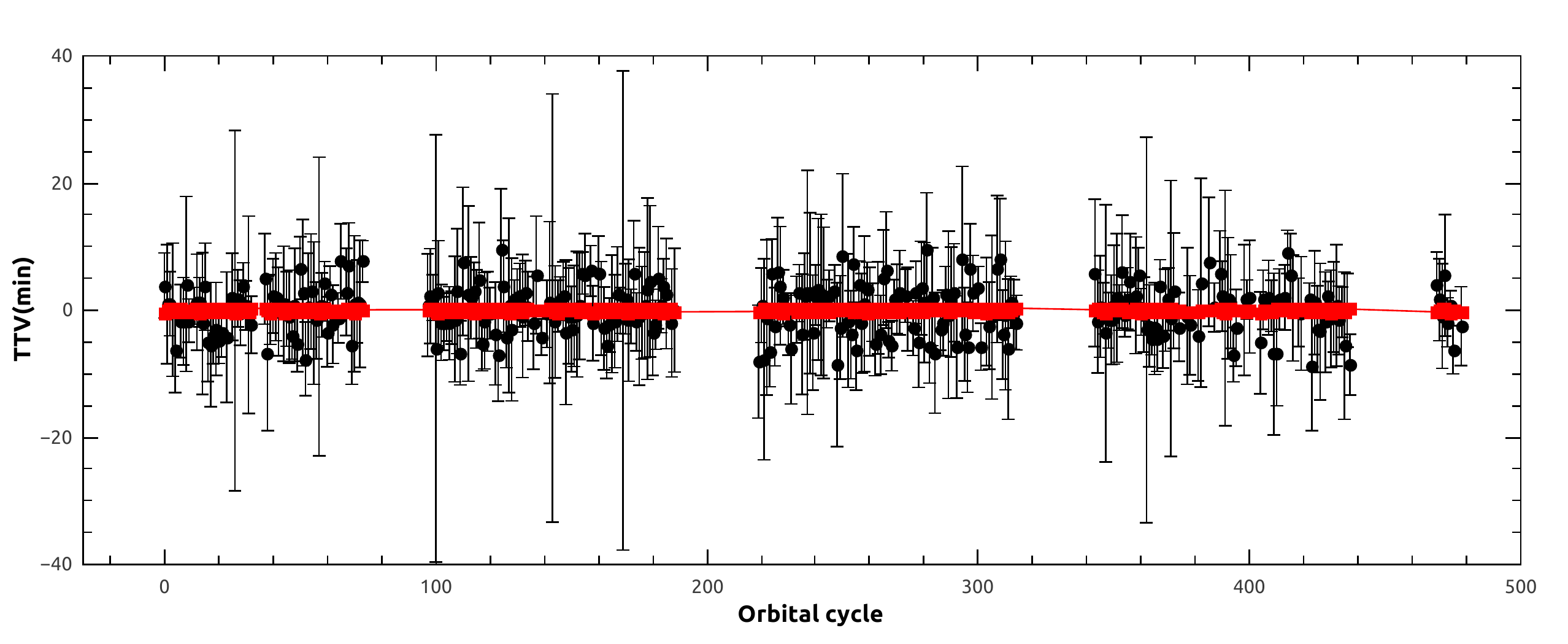}
   \caption{Optimal fitting result of Kepler-411b TTVs.}
              \label{TTVb}%
    \end{figure}

\begin{figure}
   \centering
   \includegraphics[width=9.5cm]{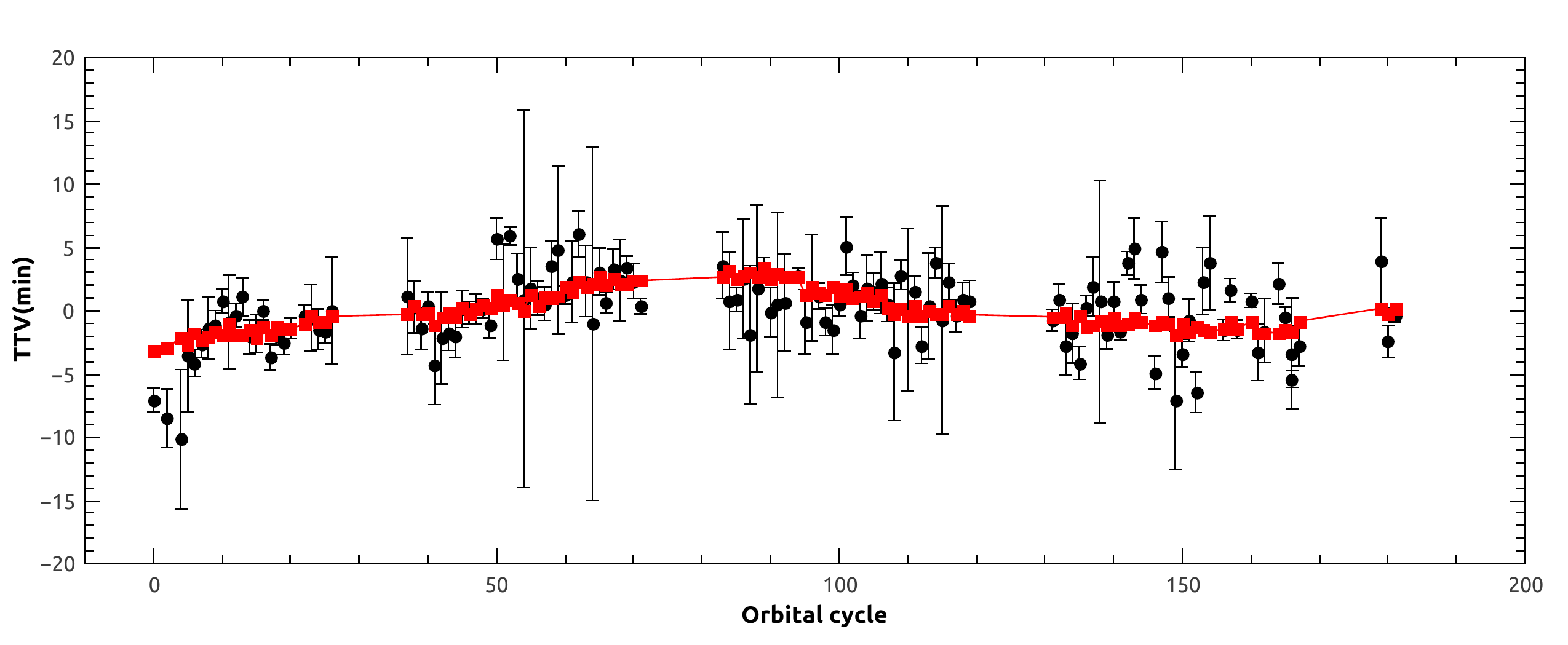}
   \caption{Optimal fitting result of Kepler-411c TTVs.}
              \label{TTVc}%
    \end{figure}

\begin{figure}
   \centering
   \includegraphics[width=9.5cm]{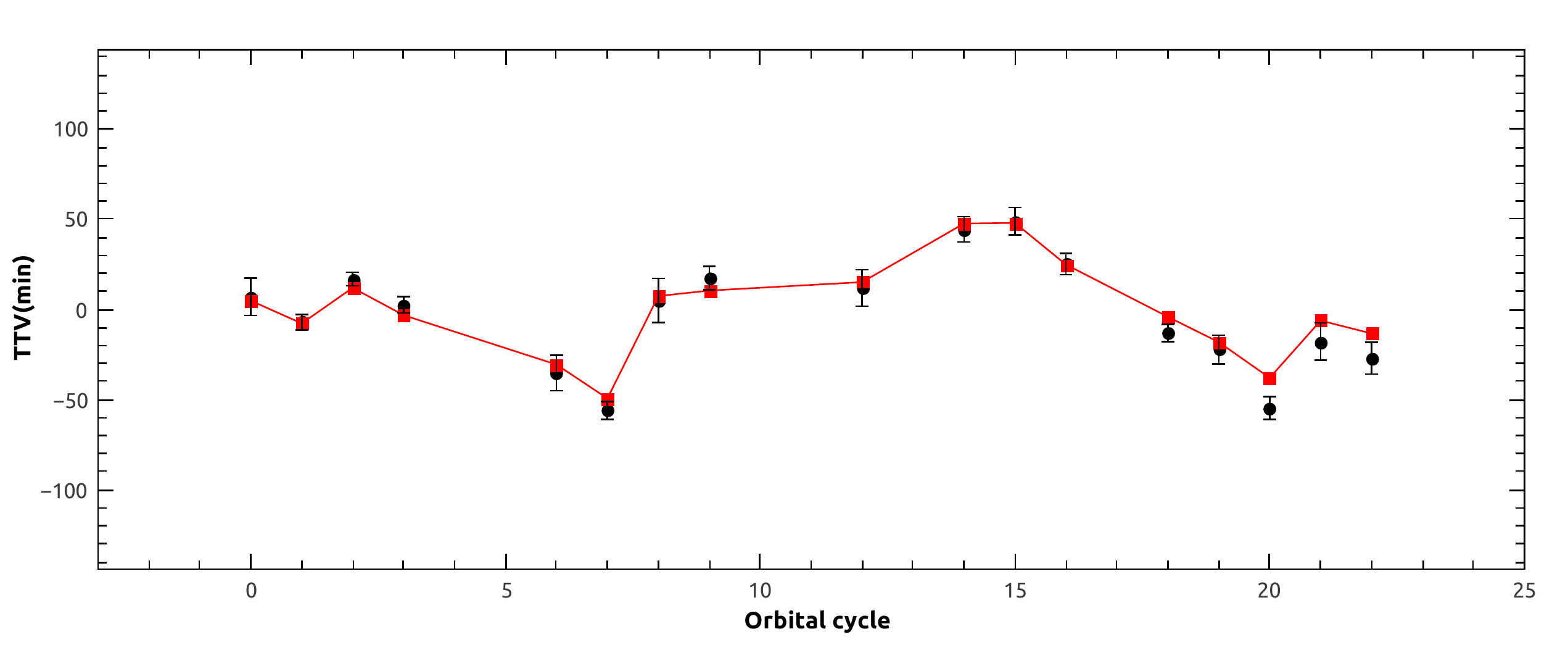}
   \caption{Optimal fitting result of Kepler-411d TTVs.}
              \label{TTVd}%
    \end{figure}

We adopt the same parameter priors and boundaries as used in Sect.~\ref{sec_initial}, except for the orbital period and inclination, since these parameters are well-constrained by transit light curves. However, the values derived from modeling the transit light curves are the averages of these parameters at the observation baseline, but the input parameters of TTVFast are a set of instantaneous variables representing the dynamical state of a system at the initial reference time~\citep{Deck2014}. Therefore, we sample the initial inclination corresponding to the nominal value and uncertainty derived from modeling transit light curves. The search interval of the orbital period is centered on the values derived from modeling the transit light curves, and is uniformly spread over intervals of 0.2~days. A typical uncertainty of 0.2~days is much larger than the uncertainty of orbital period derived from modeling transit light curves, but this value likely covers the possible variation of instantaneous period.  The optimal physical parameters resulting from the numerous TTV inversions are illustrated in Table~\ref{table:1} for the four-planet orbital architecture. These optimal models are plotted in Figs.~\ref{TTVb}, ~\ref{TTVc}, and ~\ref{TTVd} with the measured TTV signals.

Many authors have shown the existence of degeneracies between planet mass and eccentricity, and even the differential argument of orbital periastron of two planets (i.e., $\Delta \omega=|\omega_1-\omega_2|$) extracted from TTVs near by first-order (or higher-order)
mean-motion resonances~\citep{Boue2012,Lith2012,Deck2015,Gozd2016,Macd2016,Nesv2016,Agol2017}. These imply that the parameter spaces of the TTV inversion are highly correlated and hence insufficient samplings can easily misrepresent the measured TTVs. In order to validate our TTV inversion results, we carry out  additional TTV modeling with two different eccentricity priors for four-planet orbital architectures: (i) an eccentricity fixed to zero, and (ii) eccentricities drawn from a uniform distribution $U(0, 0.1)$. When the orbits of all (or a subset of the) planets in the Kepler-411 system are circular, the measured TTVs are not well reproduced. In the second case, all  TTVs are well fitted, and the results are comparable to the case when the modeling of the eccentricities is restricted to $U(0, 0.2)$. As expected, the optimum fits for the planetary masses are mutually consistent at the 1$\sigma$ significance level for the two different eccentricity priors, but the optimum fits for the orbital eccentricities are different. Similar results were also found in the TTV inversions of \citet{Macd2016} for the Kepler-80 system. Therefore, we are confident that the masses extracted from our TTV modeling are reliable global optimal values.

\subsection{Dynamical stability}

\subsubsection{Stability analyses with analytic criterions }

As described in Sect.~\ref{inversion_code}, the TTV inversion code includes a Hill stability criterion for two-planet systems. Although this initial check filters out the least stable systems, it cannot guarantee the long-term stability of the derived orbital architecture of a four-planet system. On one hand, there is no analytical criterion for assessing the long-term stability of multi-planet systems (>2 planets) like that  for two-planet systems~\citep{Glad1993,Cham1996}. \citet{Cham1996}, \citet{Smit2009}, \citet{Liss2011b}, and similar numerical studies have found that long-term stability of multi-planet systems typically requires the mutual separations between planets to be at least ten mutual Hill radii, which is much larger than the cautious limit of $2\sqrt{3}$ mutual Hill radii required by the Hill stability criterion that is implemented in our TTV inversion code~\citep{Glad1993,Cham1996}. Meanwhile, these criteria are only valid under the assumptions of low mutual inclinations and small eccentricities. These two restrictions are usually not well-defined; limiting values of $1^\circ-2^\circ$ and $e=0.1-0.2$, respectively, are often adopted~\citep{Petr2015,Macd2016}. Apparently, these restrictions and stability criteria are well satisfied by the nominal orbital architecture of the Kepler-411 system inverted from the TTV data. On the other hand, the Hill criterion provides no information about the long-term behavior of the system, and repeated weak interactions between planets in Hill stable orbits may still lead to ejections and/or physical collisions; these  are referred to as Lagrange unstable~\citep{Petr2015}. The chaotic orbits that are generated primarily by first-order (or higher-order) resonance overlap are eventually subjected to large-scale variation of the semi-major axes and hence become Lagrange unstable. Based on previous studies on the chaotic behavior induced by the first-order resonance overlap (e.g.,~\citet{Wisd1980,Dunc1989}), \citet{Deck2013} supply the condition that leads to chaotic behavior in a two-planet system (see Eq.~50 in~\citet{Deck2013}). We apply this criterion to estimate the nominal orbital architecture extracted from the TTV data of the Kepler-411 system, and find that the orbital configuration is far away from the chaotic motion. Therefore, we conclude that the nominal orbital architecture of the Kepler-411 system extracted from measured TTVs satisfies both the Hill and Lagrange stability criteria.

\subsubsection{Stability analysis using numerical dynamical simulation}

After completing the inversion of TTVs, we obtain the configuration of the Kepler-411 system at the reference time (i.e., $BJD_{TDB}-2454833=130$). Using these as initial conditions, we carry out simulations to evaluate the long-term dynamics of the system. TTVFast is useful for carrying out short-term simulations of planetary systems, such as for the baseline of the Kepler observations (i.e., several years), while general relativity and tidal effects are not included in the code~\citep{Deck2014}. These effects, however, can become important on timescales beyond 1~Myr. 

In order to meet our requirements, we employ another publicly available $N$-body package REBOUND, which uses the Wisdom-Holman symplectic integrator WHFast~\citep{Rein2012,Rein2015,Rein2017}, to evolve the system for 1~Myr. We note that these integrations are not intended to provide a comprehensive overview of the dynamics of the planetary system, but are merely used to check whether the system can remain stable for at least 1 Myr. In these simulations, the stellar mass is fixed to 0.87~$M_\odot$, and the integration step is set to 0.15~days (roughly 0.05 times the orbital period of the innermost planet). We obtain the initial semi-major axis  for each planet from its period using Kepler's third law. With the exception of orbital semi-major axis, the initial conditions for the other Jacobi orbital elements, such as the eccentricity $e$, the inclination $i$, the longitude of ascending node $\Omega$, the argument of periastron $\omega,$ and the mean anomaly $M$, are fixed to the values derived from the TTV inversions so as to improve the efficiency of the simulations. This choice is made because the primary determinants are the mutual distances of the planets for the long-term stability of the system with low eccentricities (<0.2) and mutual inclinations (< 2$^\circ$). We draw 100 initial values for the semi-major axis from a normal distribution $N(a_{0}, \sigma_{a})$, where $a_{0}$ represents the nominal value obtained from the TTV inversion, and $\sigma_{a}$ the spread in the semi-major axis, which is estimated using Monte Carlo simulations which include the uncertainties in the mass of the host star, the planetary masses, and the orbital periods. 

We classify stable planetary systems as those in which the minimum distance between the planets never violates the Hill stability criterion during the simulations, since the systems that violate the criterion tend to be short-lived due to close planet-planet encounters~\citep{Glad1993}. In addition, we also require the variations in the semi-major axes to be smaller than 10\% in cases where a planet is ejected from the system or collisions occur with the star. Among all simulations, the majority ($\sim 60\%$) remain stable beyond 1~Myr. 

\section{Discussion}

\subsection{The possibility that Kepler-411 e is on a very eccentric and inclined orbit }
\label{inclined orbit}

The constraint of mutual inclination in Section \ref{sec_initial} mainly comes from considerations of the long-term stability and less-significant TDV signals of the Kepler-411 system if we hypothesize that the system is similar to most Kepler compact multi-planet systems. \citet{Nesv2009} finds that orbits with mutual inclinations $\ge 30^\circ$ are violently unstable in the most compact planetary systems. For the Kepler-108 system,  the mutual inclination of roughly $ 24^\circ$ triggers significant TTV and TDV signals on the near 4:1 orbital architectures~\citep{Mill2017}. Therefore, in order to efficiently implement the TTV inversion of Kepler-411 d, we explicitly set the upper limit of mutual inclination to $30^\circ$. However, if the perturber is on a wider orbit that is different from most Kepler compact orbital architecture, this kind of orbital architecture, even with larger eccentricity and mutual inclination, will probably be stable in the long term and will likely not trigger large TDV signals on the baseline of primary Kepler observations. In addition, this kind of orbital architecture can also probably induce large TTV signals on Kepler-411 d, due the fact that the mechanism involved is different from the MMR;  for example, as the distance of the perturber changes due to its eccentricity, the orbital period of the perturbed planet changes. (see \citet{Agol2005}). Most importantly, the period of TTVs induced by a perturber on a very eccentric and wide orbit is approximately equal to the orbital period of the perturber~\citep{Agol2005}.

We study the possibility that Kepler-411 e is on a very eccentric and inclined orbit. Firstly, we analyze the period of the TTVs of Kepler-411 d using the periodogram tool and find that the power of the periodogram exhibits an apparent peak on the period of about $918\pm1.5$ days. This period should be very close to the orbital period of the perturber if the TTVs of Kepler-411 d are induced by a massive perturber with an eccentric, wide-orbit architecture.  Secondly, we set the upper limits of eccentricity and mutual inclination of the perturber to 0.9 and $30^\circ$, respectively; the  upper limit of the eccentricity of Kepler-411 d  is set to 0.5, however, in order to maintain the long-term stability of the inner three planets. In addition, we search for the orbital period of the  perturber in the interval between 908  and 928 days, and for the masses of both Kepler-411d and the perturber  between 0.1 $M_\oplus$ and 15$M_{Jup}$.  Finally, we find that this kind of orbital architecture cannot accurately reproduce the TTVs of Kepler-411d. Therefore, we confirm that the orbital architecture of $P_{pert}=31.5$ days is the optimal solution for the TTVs of Kepler-411d.

\subsection{Possible influence of the stellar companion of Kepler-411}
\label{Possible_influence}

Although the planetary nature of Kepler-411b and Kepler-411c have been confirmed by~\citet{Mort2016} and~\citet{Wang2014}, the status of the third candidate, Kepler-411d, is still ambiguous. Adaptive optics and speckle imaging observations indicate a low false-positive rate for Kepler-411d, and also suggest that no other stellar companion is present, in addition to the one already known~\citep{Law2014,Wang2015,Furl2017}. 
In this section we consider the possibility that Kepler-411d orbits the companion, instead of Kepler-411 itself. The Kepler-band differential magnitude between Kepler-411 and its companion is roughly 3.0~mag, derived using the values and empirical relations of~\citet{Wang2015} and~\citet{Furl2017}. We convert this differential magnitude into a flux ratio $f_A/f_B$=15.85, where $f_A$ and $f_B$ denote the fluxes of Kepler-411 and its stellar companion, respectively. 

To test this hypothesis, we assume that the   transit signals of Kepler-411d originate from the companion, which is eclipsed by another body. These hypthetical eclipse signals are diluted by the flux of Kepler-411. We employed STMT to fit the short-cadence transit light curves of Kepler-411d, fixing the dilution factor to 15.85 (i.e., the flux ratio). The $\chi^2$ of best fit is 20369. Subsequently, we adopt the same procedure to fit the same data, but now fixing the dilution factor to 1/15.85. We note that the dilution of the companion to the transit light curves of Kepler-411 is negligible in this case. The $\chi^2$ is now 20286, under the assumption that Kepler-411d orbits Kepler-411. The second hypothesis leads to a much better fit to the data than the first. Finally, we adopt the Bayesian information criterion (BIC)~\citep{Schw1987}, proven to be very useful for model selection, to compare both hypotheses. We derive a $\Delta{BIC}$ of 83, which strongly supports the second hypothesis. This means that based on the Bayesian statistics
theory (i.e., approximately estimated by using $\exp\{-\Delta{\rm BIC}/2.0\}$) the false-positive probability of the hypothesis that Kepler-411d orbits Kepler-411 is about $9.4\times 10^{-19}$ . 

We have demonstrated that the addition of planet Kepler-411 e is indispensable to reproduce the measured TTVs of Kepler-411 d. Furthermore, the orbital period ratio of Kepler-411 c and Kepler-411 e is in the vicinity of four, which implies that there is a chance of detecting the influence of Kepler-411 e  in the TTVs of Kepler-411 c if the two planets belong to the same system. We compare the fit results of the TTVs of Kepler-411 b and Kepler-411 c  based on a two-planet architecture and a four-planet one, respectively. For the two-planet architecture, the minimum $\chi^2_{\nu}$ of the TTV fitting of Kepler-411 b and Kepler-411 c is equal to 972. For the four-planet system,  the minimum $\chi^2_{\nu}$ of the TTV fittings of both planets decreases to 854 by considering the gravitational perturbations of Kepler-411 e to Kepler-411 c.  We still use the BIC to complete the model selection between these two orbital configurations. The BIC value for the two-planet model is much larger than that for the four-planet one (i.e., $\Delta BIC=81$); such a large $\Delta BIC$ is in favor of both planets d and e orbiting Kepler-411.

The presence of a wide-orbit stellar companion around Kepler-411 may produce TTV signals and can influence the dynamical stability of the Kepler-411 planetary system. Here, we discuss these possibilities. The possibility of the large TTVs of Kepler-411d being due to the stellar companion can easily be ruled out, based on three different TTV amplitudes of the  transiting planets of Kepler-411. The TTV signals induced by the  stellar companion of Kepler-411 are primarily due to the light-travel effect, that is, the time delay or advance owing to the reflex motion of the Kepler-411 system induced by the companion~\citep{Irwi1952,Agol2005,Bork2016}. Therefore, all the planetary companions of Kepler-411 should exhibit relatively large TTVs, and we have not considered the stellar companion as the potential perturber during our TTV inversion.

The gravitational perturbations on the Kepler-411 planetary system due to the wide-orbit companion are negligible, and do not influence their dynamical stability. \citet{Wang2015} estimated the separation of the companion to Kepler-411 to be $\sim 600$~AU, and the mass of the stellar companion to be $\sim0.13M_{\odot}$ assuming physical association. The corresponding orbital period of the companion is then roughly $15\ 000$~years, if we assume that the semi-major axis is equal to the projected separation. The gravitational perturbation to the Kepler-411 system is therefore most likely negligible and should not significantly change the results of long-term dynamical simulations even if the stellar companion is actually physically related to the Kepler-411 system. Therefore, we have not included the stellar companion in long-term dynamical simulations.

\subsection{Constraints on the compositions of the' transiting
planets of Kepler-411}

Before analyzing the bulk compositions, we estimate the age of the Kepler-411 system according to the gyrochronological relation of \citet{Barn2007} (see their Eq.~2). We estimate an age of $212\pm31$~Myr based on the color index, the rotation period of Kepler-411, and the systematic uncertainty in the relation. We can therefore make the reasonable assumption that the gyrochronological age of the host star Kepler-411 is equal to that of the planetary system. 

In addition to the age, the incident stellar flux also affects the radii of gaseous and H/He-enriched planets according to the interior structure and thermal evolution models of~\citet{Fort2007} and \citet{Lope2014}. In order to directly compare with these models, we derive the scaled orbital separations and incident fluxes of the  transiting planets of Kepler-411 relative to those of the Earth, following the expressions in \citet{Fort2007}. The respective scaled separations for Kepler-411b$-$d are $0.048\pm0.004$ AU, $0.095\pm0.008$ AU, and $0.358\pm0.029$AU,  and the corresponding incident fluxes are $412\pm71$, $114\pm18$, and $7.9\pm1.3$ times what the present-day Earth receives from the Sun. 

\begin{figure}
   \centering
   \includegraphics[width=9.5cm]{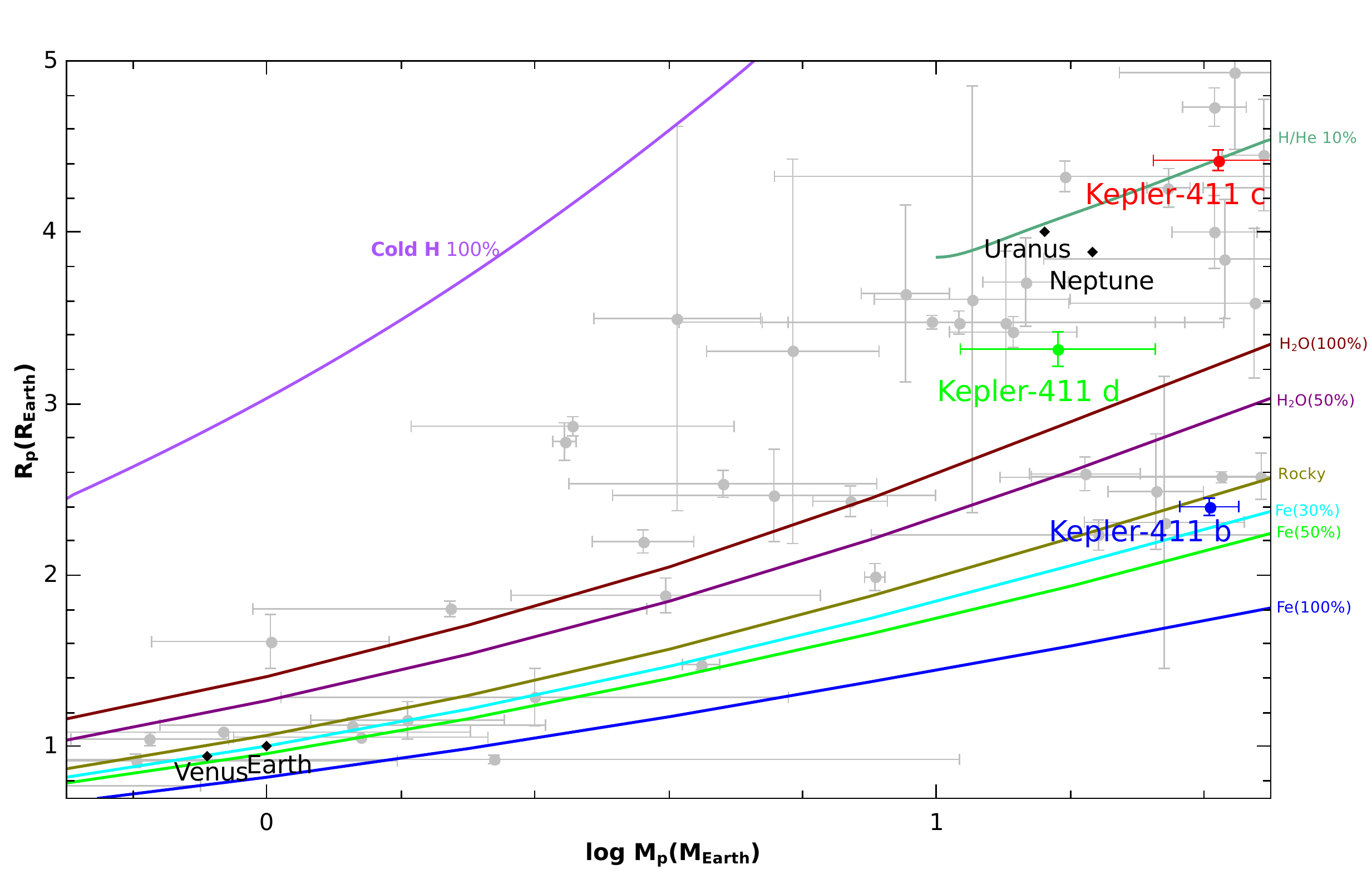}
   \caption{Mass-radius diagram of the  transiting planets of Kepler-411, including other exoplanets with known masses and radii, obtained from Extrasolar Planets Encyclopaedia. Theoretical models for different planetary compositions are indicated with color curves.}
              \label{mass-radius}%
    \end{figure}

We show the mass-radius diagram in Fig.~\ref{mass-radius}, which includes the transiting planets of Kepler-411 and some transiting planets with precisely measured masses and radii obtained from The Extrasolar Planets Encyclopaedia\footnote{www.exoplanet.eu}. The model of solid planets (i.e., pure-iron, rocky, and water/ice planet) is obtained from Table~2 of \citet{Zeng2016}. The cold-hydrogen model, and the 10\% H/He envelope model are obtained from \citet{Seag2007} and \citet{Fort2007}, respectively. According to the expression in \citet{Zeng2016}, we find that the sub-Neptune-sized planet Kepler-411b contains a $21\pm21$\% mass fraction of iron core with a rocky mantle. 

For Kepler-411c and Kepler-411d, however, the analysis is more complicated than that of the solid planets. This is because the radius of a H/He-enriched planet depends on many variables, such as age, incident stellar flux, the H/He envelope mass fraction and even the composition of the heavy-element core~\citep{Fort2007,Lope2014}. Nevertheless, \citet{Lope2014} show that the H/He envelope fraction plays a crucial role in determining the radius of a sub-Neptune planet. On the mass--radius diagram (see Fig.~\ref{mass-radius}), we find that Kepler-411c is adjacent to the region of planets with a 10\% H/He envelope. Also, the two solar-system ice giants (Uranus and Neptune) are located near the position related to this model. Therefore, just like Uranus and Neptune, Kepler-411c should have a low-density H/He envelope in order to be consistent with our mass and radius measurements, but with most of the mass of the planet in the dense core. 

However, the model with a 10\% H/He envelope is calculated under the assumption that a planet contains a 50/50 ice-rock core and receives the incident flux identical to that of the present-day Earth, and these assumptions are not applicable to Kepler-411c. In order to derive an appropriate model for Kepler-411c and Kepler-411d, we linearly interpolate the model radii of the H/He-enriched planets with a $25 M_\oplus$ and $10 M_\oplus$ ice/rock core at an age of 300 Myr, respectively, in Table~2 of \citet{Fort2007}, whose irradiations are 100 times higher than that of the present-day Earth. We derive rock-ice cores with masses of $24.2\pm4.8 M_\oplus$ and $14.4\pm4.4 M_\oplus$ for Kepler-411c and Kepler-411d, respectively, which corresponds to H/He envelope fractions of $8.3\%$ and $5.2\%$. Using the models of \citet{Lope2014}, we find that if the interior has an Earth-like composition with an iron core and a rocky mantle, the H/He envelopes around Kepler-411c and Kepler-411d should have mass fractions of $4.7\%$ and $1.7$\%, respectively. 

\section{Summary}

We investigate the orbital and physical properties of the Kepler-411 planetary system. We combine Kepler photometry data and new TTV measurements from all Kepler quarters with previous adaptive optics, speckle imaging, and dynamical simulation results. Based on the initial conditions extracted from measured TTV signals, we carry out a series of long-term dynamical simulations and find that most simulations remain stable beyond 1~Myr. Our finding can be summarized as follows.
\begin{enumerate} 
\item We obtain a mass of $25.6\pm2.6\,M_\oplus$ for Kepler-411b and a mass of $26.4\pm5.9\,M_\oplus$ for Kepler-411c. 
\item We confirm the planetary nature of the previously known exoplanet candidate KOI1781.03, now known as Kepler-411d, and find its mass to be $15.2\pm5.1\,M_\oplus$. 
\item By assuming a near-coplanar system (mutual inclination restricted $\lesssim 30^\circ$), we have also discovered a nontransiting exoplanet Kepler-411e with a mass of $10.8\pm1.1\,M_\oplus$ on a 31.5-day orbit. This planet shows evidence of dynamical interaction with Kepler-411d.
\item With the newly measured masses and radii, we confirm that Kepler-411c and Kepler-411d belong to the group of planets with a massive core and a significant fraction of volatiles. Although Kepler-411b possesses a sub-Neptune size, it belongs to the group of rocky planets. 
\end{enumerate}

Kepler-411 is a slight bright host star ($V = 12.3$~mag), which makes the planetary system a promising target for follow-up spectroscopic observations. The RV semi-amplitudes of the host star reach up to 12\,m\,s$^{-1}$ and 9\,m\,s$^{-1}$ respectively induced by Kepler-411b and c, and both are on short-period orbits ($P_{orb}$<10 days). The RV measurements, which
are independent of the TTV technique, can provide more constraints on both masses and orbital eccentricities of the planets of Kepler-411.  Although the presence of strong magnetic activity of the host star makes the characterization of the Kepler-411 system using the RV technique less feasible, the TESS mission provides an excellent opportunity to resolve this problem; the TESS mission will supply  continuous, high-precision photometric data for the Kepler-411 system covering at least 27 days. If sufficient high-precision RV measurements ($\sigma_{RV}\lesssim 1$\,m\,s$^{-1}$) were obtained with TESS photometery simultaneously by employing a state-of-the-art high-precision velocimeter, such as HARPS or CARMENES, and so on, the stellar activity-induced RV variations could be appropriately modeled; there is also a chance that the RV signal of the newly discovered nontransiting planet Kepler-411e (i.e., the RV semi-amplitude $K_e=2\,$m\,s$^{-1}$) could be detected \citep{Tuom2014,Dumu2015,Barn2017,Jone2017,Osha2018}.

\begin{acknowledgements}

We thank S. Czesla and S. Jia for helpful discussions before preparing the manuscript. We also thank the anonymous referee for valuable suggestions and comments. This research has made use of the Mikulski Archive for Space Telescopes, which is a NASA funded project to support and provide to the astronomical community a variety of astronomical data archives. This research also has made use of the Keck Observatory Archive (KOA), which is operated by the W. M. Keck Observatory and the NASA Exoplanet Science Institute (NExScI), under contract with the National Aeronautics and Space Administration. This work is supported by National Natural Science Foundation of China through grants No. U1531121, No. 10873031 and No.11473066. Here, we also thank the support from China Scholarship Council (CSC) and Deutscher Akademischer Austausch Dienst (DAAD) to our work. The joint research project between Yunnan Observatories and Hamburg Observatoy is funded by Sino-German Center for Research Promotion (GZ1419). M.B.N.K. was supported by the National Natural Science Foundation of China (grants 11010237, 11050110414, 11173004, and 11573004). This research was supported by the Research Development Fund (grant RDF-16-01-16) of Xi'an Jiaotong-Liverpool University (XJTLU).

\end{acknowledgements}

%-------------------------------------- Two column figure (place early!)

\begin{table*}
\caption{Physical parameters of the planetary companions of Kepler-411.}             
\label{table:1}      
\centering          
\begin{tabular}{c c c c l l l }     % 7 columns 
\hline\hline                           
 Planet&Parameter & Value & Note         \\
  \hline\noalign{\smallskip}
$Kepler-411 b$ & $T_{0}(BJD_{TDB}-2454833)$ &302.3874$\pm0.0003$  &  Transit epoch               \\
& $P(d)$          &3.005156$\pm0.000002$  &Orbital period                  \\
&$R_{p}/R_{*}$      &0.0266$\pm0.0002$ &Planet/star radii ratio  \\
&$(R_{p}+R_{*})/a$     &$0.1055\pm0.0006 $ &Scaled total radii ratio   \\
&$b$             &$0.574^{+0.012}_{-0.015}$ & Impact parameter      \\
&$i(^\circ)$          &$87.4\pm0.1$ & Orbital inclination          \\
&$a(AU)$              &$0.0375\pm0.0008$ & Orbital separation            \\
&$R_p(R_{\oplus})$       &$2.401\pm0.053$ & Planet radius                      \\
&$T_{eq}(K)$            &$1138\pm17$   & Planet temperature     \\
&$e$                  &$0.146^{+0.005}_{-0.004}$& Orbital eccentricity     \\
&$\omega(^\circ)$   & $41.5^{+3.8}_{-3.1}$ & Argument of periastron \\
&$M(^\circ)$                      &$171.6^{+12.6}_{-5.6}$ & Mean anomaly     \\
&$M_p(M_{\oplus})$      &$25.6\pm2.6$           & Planet mass      \\
&$\rho_p(\rho_{\oplus})$   &$1.86\pm0.23$      &Planet mean density \\
\hline\noalign{\smallskip}
$Kepler-411 c$ &$T_{0}(BJD_{TDB}-2454833)$ &135.2224$\pm0.0002$  &  Transit epoch               \\
&$P(d)$          &7.834435$\pm0.000002$  &Orbital period                   \\
&$R_{p}/R_{*}$      &0.0449$\pm0.0001$ &Planet/star radii ratio  \\
&$(R_{p}+R_{*})/a$     &$0.0592\pm0.0003 $ &Scaled total radii ratio         \\
&$i(^\circ)$          &$88.61\pm0.04$ & Orbital inclination          \\
&$b$             &$0.620^{+0.007}_{-0.009}$ & Impact parameter      \\
&$a(AU)$              &$0.0739\pm0.001$ & Orbital separation            \\
&$R_p(R_{\oplus})$       &$4.421\pm0.062$ & Planet radius                      \\
&$T_{eq}(K)$            &$838\pm10$   & Planet temperature     \\
&$e$                        &$0.108^{+0.003}_{-0.004}$ &Orbital eccentricity \\
&$\omega(^\circ)$    & $103.0^{+1.6}_{-1.7}$      & Argument of periastron                  \\
&$M(^\circ)$                         &$280.5^{+4.7}_{-3.8}$        &Mean anomaly \\
&$M_p(M_{\oplus})$      &$ 26.4 \pm5.9$                 & Planet mass      \\ 
&$\rho_p(\rho_{\oplus})$  &$0.31\pm0.07$               &Planet mean density \\
\hline\noalign{\smallskip}
$Kepler-411 d$ & $T_{0}(BJD_{TDB}-2454833)$ &151.8484$\pm0.0061$  &  Transit epoch               \\
&$P(d)$          &58.02035$\pm0.00056$  &Orbital period                   \\
&$R_{p}/R_{*}$      &0.0372$\pm0.0003$ &Planet/star  radii  ratio  \\
&$(R_{p}+R_{*})/a$     &$0.0141\pm0.0003 $ &Scaled total radii ratio         \\
&$i(^\circ)$          &$89.43\pm0.02$ & Orbital inclination          \\
&$b$             &$0.7193^{+0.014}_{-0.012}$ & Impact parameter      \\
&$a(AU)$              &$0.279\pm0.004$ & Orbital separation            \\
&$R_p(R_{\oplus})$       &$3.319\pm0.104$ & Planet radius                      \\
&$T_{eq}(K)$            &$410\pm10$   & Planet temperature     \\
&$e$                         &$0.128^{+0.003}_{-0.003}$   &Orbital eccentricity \\
&$\omega(\circ)$    & $30.5\pm1.1$                      & Argument of periastron                                   \\
&$M(^\circ)$                        &$150.5\pm1.1$                      &Mean anomaly  \\
&$M_p(M_{\oplus})$   &$15.2\pm5.1$                       & Planet mass       \\
&$\rho_p(\rho_{\oplus})$  &$0.42\pm0.15$                &Planet mean density      \\
\hline\noalign{\smallskip}
$Kepler-411 e$ & $P(d)$          &31.509728$\pm0.000085$  &Orbital period                     \\
&$i(^\circ)$          &$88.04\pm0.02$ & Orbital inclination          \\
&$b$             &$1.688^{+0.006}_{-0.006}$ & Impact parameter      \\
&$a(AU)$              &$0.186\pm0.003$ & Orbital separation            \\
&$T_{eq}(K)$            &$503\pm9$   & Planet temperature     \\
&$e$                  &$0.016^{+0.002}_{-0.001}$           &Orbital eccentricity \\
&$\omega(^\circ)$    & $-103.3\pm1.5$          & Argument of periastron                                   \\
&$M(^\circ)$                        &$161.1\pm1.9$           &Mean anomaly  \\ 
&$M_p(M_{\oplus})$      &$10.8\pm1.1$          & Planet mass       \\ 
\hline\hline                  
\end{tabular}
\end{table*}

\begin{table*}
\caption{The mid-transit times of the transit events of  Kepler-411 d. This is an excerpt from the
  complete table that is available at the CDS, provided here as a guide to its format and content.}             
\label{table:2}              
\begin{tabular}{lllllllrrrrr}     % 7 columns 
\hline\hline                           
 Orbital cycle & $ BJD_{TDB}-2454833$ & Uncertainty & (O-C) (d)  &  Orbital cycle & $ BJD_{TDB}-2454833$ & Uncertainty &   (O-C) (d)\\
  \hline\noalign{\smallskip}
0       &       151.849785      &       0.004784        &       0.001927        &       14      &       964.161382      &       0.003244        &       0.034233        \\
1       &       209.860345      &       0.001950        &       -0.007463       &       15\tablefootmark{f}     &       1022.184991     &       0.003526        &       0.037892        \\
2       &       267.897426      &       0.001746        &       0.009669        &       16\tablefootmark{ft}   &       1080.188858     &       0.002728        &       0.021810        \\
3       &       325.907917      &       0.002104        &       0.000211        &       18\tablefootmark{f}     &       1196.203068     &       0.002248        &       -0.003879       \\
6\tablefootmark{o}      &       499.942828      &       0.004534        &       -0.024726       &       19\tablefootmark{ft}   &       1254.217127     &       0.003668        &       -0.009769       \\
7\tablefootmark{d}      &       557.948817      &       0.002300        &       -0.038686       &       20\tablefootmark{fd}   &       1312.215050     &       0.002916        &       -0.031795       \\
8       &       616.011568      &       0.005654        &       0.004115        &       21\tablefootmark{fd}   &       1370.260992     &       0.004790        &       -0.005803       \\
9       &       674.040629      &       0.002984        &       0.013227        &       22      &       1428.275029     &       0.004068        &       -0.011715       \\
\hline\hline                  
\end{tabular}
\tablefoot{ d shows that this transit is a double-transit event. f  shows that this transit is selected to construct the fiducial model. o shows that one-spot model is used when the transit time is measured. t shows that two-spot model is used when the transit time is measured.\\
}
\end{table*}

\begin{table*}
\caption{The mid-transit times of the transit events of  Kepler-411 b. This is an excerpt from the
  complete table that is available at the CDS, provided here as a guide to its format and content.}             
\label{table:3}              
\begin{tabular}{lllllllrrrrr}     % 7 columns 
\hline\hline                           
 Orbital cycle & $ BJD_{TDB}-2454833$ & Uncertainty & (O-C) (d)  &  Orbital cycle & $ BJD_{TDB}-2454833$ & Uncertainty &   (O-C) (d)\\
  \hline\noalign{\smallskip}
0       &       152.132025      &       0.002370        &       0.002571        &       71      &       365.496323      &       0.003010        &       0.000870        \\
1       &       155.135147      &       0.004338        &       0.000538        &       72      &       368.501186      &       0.004628        &       0.000579        \\
2       &       158.140339      &       0.002332        &       0.000576        &       73      &       371.511010      &       0.001506        &       0.005247        \\
3       &       161.144853      &       0.004848        &       -0.000066       &       97      &       443.629959      &       0.003742        &       0.000478        \\
4       &       164.145569      &       0.003076        &       -0.004504       &       98      &       446.636112      &       0.003452        &       0.001476        \\
6       &       170.159068      &       0.002658        &       -0.001315       &       99      &       449.641084      &       0.001664        &       0.001293        \\
7       &       173.164217      &       0.003188        &       -0.001321       &       100     &       452.640675      &       0.015562        &       -0.004271       \\
8       &       176.173433      &       0.006372        &       0.002740        &       101     &       455.651887      &       0.003828        &       0.001786        \\
9       &       179.174597      &       0.001572        &       -0.001251       &       102     &       458.653834      &       0.002770        &       -0.001421       \\
11      &       185.186523      &       0.001348        &       0.000366        &       103     &       461.658870      &       0.001958        &       -0.001541       \\
\hline\hline                  
\end{tabular}
\tablefoot{Table marks (i.e., d,f,o,t) in this table have same meaning with Table 2(see the notes of Table 2 for details).}
\end{table*}

\begin{table*}
\caption{The mid-transit times of the transit events of  Kepler-411 c. This is an excerpt from the
  complete table that is available at the CDS, provided here as a guide to its format and content.}             
\label{table:4}              
\begin{tabular}{lllllllrrrrr}     % 7 columns 
\hline\hline                           
 Orbital cycle & $ BJD_{TDB}-2454833$ & Uncertainty & (O-C) (d)  &  Orbital cycle & $ BJD_{TDB}-2454833$ & Uncertainty &   (O-C) (d)\\
  \hline\noalign{\smallskip}
0       &       158.721100      &       0.000438        &       -0.004496       &       84      &       816.818490      &       0.001792        &       0.000572        \\
2\tablefootmark{t}      &       174.388924      &       0.001072        &       -0.005537       &       85      &       824.652971      &       0.000436        &       0.000621        \\
4       &       190.056635      &       0.002548        &       -0.006691       &       86      &       832.488566      &       0.002192        &       0.001783        \\
5\tablefootmark{o}      &       197.895637      &       0.002040        &       -0.002122       &       87\tablefootmark{d}     &       840.319902      &       0.002534        &       -0.001313       \\
6       &       205.729685      &       0.000494        &       -0.002506       &       88\tablefootmark{d}     &       848.156860      &       0.003062        &       0.001213        \\
7       &       213.565164      &       0.000516        &       -0.001459       &       89\tablefootmark{t}     &       855.992499      &       0.000334        &       0.002419        \\
8       &       221.400437      &       0.001136        &       -0.000618       &       90      &       863.824415      &       0.000900        &       -0.000097       \\
9       &       229.235080      &       0.000504        &       -0.000408       &       91\tablefootmark{t}     &       871.659257      &       0.003394        &       0.000312        \\
10      &       237.070789      &       0.000424        &       0.000869        &       92\tablefootmark{d}     &       879.493829      &       0.001780        &       0.000452        \\
11      &       244.904073      &       0.001712        &       -0.000280       &       94      &       895.164195      &       0.000262        &       0.001953        \\ 
\hline\hline                  
\end{tabular}
\tablefoot{Table marks (i.e., d,f,o,t) in this table have same meaning with Table 2(see the notes of Table 2 for details).}
\end{table*}


\end{document}